# Microparticles manipulation with a three-dimensional closed optical trap

Chunyu Zhang[1,*], Xinran Fang[1], Pengcheng Mao[2], Dongxu Zhao[3,*]

## Abstract

Gaussian optical tweezers have strong phototoxicity to bioactive substances and it is difficult to achieve capture and manipulation of multiple particles. Moreover, vortex optical tweezers face several challenges, such as weak axial confinement and dependence on complex optical elements like a spatial light modulator (SLM). Therefore, we design and construct a novel three-dimensional (3D) "optical spindle trap" (OST) system that does not require a SLM. By employing intracavity mode modulation and a simple extra-cavity lens modulation, we generate a fully enclosed 3D dark potential well with zero central intensity. Experimental results demonstrate that the system achieves stable trapping of single or multiple micrometer-sized particles with excellent 3D confinement and enables the trapping of mouse hepatocytes under low-power conditions. The system exhibits high physical stability. Its closed dark-field structure can also reduce phototoxic damage to biological samples. Furthermore, the dimensions of the optical trap can be flexibly tuned to suit various application scenarios. This study provides a novel, highly efficient, gentle, and low-cost optical trap platform for biomedicine and micro/nanoscale manipulation.

[1]School of Medical Science and Engineering, Beijing Institute of Technology, Beijing 100081, China

[2]Analysis & Testing Center, Beijing Institute of Technology, Beijing 100081, China

[3]School of life science, Beijing Institute of Technology, Beijing 100081, China

[*]To whom correspondence should be addressed. E-mails: zzccyy1977@bit.edu.cn (Chunyu Zhang) zhaodx@bit.edu.cn (Dongxu Zhao)

# Introduction:

Since Arthur Ashkin first observed that light could exert forces on microscopic particles at Bell Laboratories in 1969[1], optical tweezers have become pivotal tools that propel the development of many fields and applications. Currently, optical tweezers technology has been widely used in living animals[2], nano-rotors[3,4], micro-robot manipulation[5,6], three-dimensional (3D) imaging microscopes[7,8], biomolecules[9-11], quantum science[12-14] and other research. Optical tweezers generally use Gaussian beams, focused by a high-numerical-aperture (NA) objective lens, to form an optical trap[15]. By inserting a specialized lens or a spatial light modulator (SLM) before the high-NA objective lens, the trapping field can be shaped into specific profiles[16-21]. Furthermore, some experiments adjust the circular polarization state of the incident beam to induce rotation of the captured object [22-25] or use an angular optical trap[26,27]. However, since the light intensity is distributed over a larger area, the trapping strength is low. Therefore, higher laser powers are typically required to generate sufficient optical gradient forces for stably capturing microparticles.

Optical tweezers are one of the few nano-tools that can manipulate and measure single molecules and organelles in living cells without destroying cell walls and inserting non-endogenous probes[28]. High laser intensity or prolonged exposure during optical trapping can result in phototoxicity[29,30], which severely constrains experimental feasibility and duration. To mitigate this limitation, subsequent research has transformed Gaussian beams into Laguerre-Gaussian (LG) modes to realize hollow-beam optical tweezers[18,31–33] by using computer-generated holograms (CGHs) and SLM. LG beams can confine particles within their central intensity null, which benefits primarily from scattering forces. If the particle is transparent and the refractive index is lower than the surrounding medium, the optical gradient force will establish a stable trapping potential along the beam axis. However, if the refractive index relationship is reversed, the particle will be expelled from the focal region by a Gaussian beam[34]. Conversely, it has been experimentally demonstrated that LG-based traps can stably confine low-refractive-index particles in three dimensions[35]. Alternatively, a SLM can be used to generate an optical Archimedes' screw for trapping and transporting absorbing particles[36].

Nevertheless, the phase calibration and dynamic wavefront modulation of SLMs frequently induce optical interference[37-39], diffraction[40], and polarization effects[41]. Furthermore, the nonlinear relationship between the SLM modulated phase and the driver voltage could be influenced by temperature fluctuations, device aging, and other factors[42]. Additionally, there is a contradiction between the maximum achievable phase modulation range and the device refresh rate, making it challenging to optimize both simultaneously. Ideally, when the SLM pixel pitch approaches the optical wavelength, larger diffraction angles can be achieved. However, this could cause low diffraction efficiency in practice.

Furthermore, liquid crystal on silicon (LCOS) spatial light modulators inherently absorb a portion of the incident laser energy, causing the device temperature to rise, thereby inducing optical phase instability and affecting modulation accuracy. If the device is exposed to high temperatures for a long time, it will cause permanent physical damage. Therefore, when using a pulsed laser as a light source, a large amount of heat accumulates in a very short time, which can destroy the liquid crystal layer instantly. Additionally, LCOS is not suitable for laser modulation in the ultraviolet spectrum, further limiting the usable spectral range. All these factors constrain the average laser power when using the SLM. Although we can increase this threshold by adding external cooling, this also increases the complexity and cost of the system. Moreover, precise wavefront shaping with SLM typically relies on computationally intensive phase-retrieval algorithms and intricate calibration routines. This not only escalates hardware and software expenditures but also significantly complicates system alignment and experimental optimization.

In some experiments, atomic force microscopy (AFM) is constrained by inherent limitations in force resolution and sensitivity. While magnetic tweezers can also capture and manipulate particles, they exhibit comparatively lower spatiotemporal resolution than optical tweezers. Therefore, in many biophysical applications, optical tweezers are the first choice because of their precise force and torque control, high spatiotemporal resolution, exceptional versatility, and broad dynamic range[43].

In this study, we developed an optical trapping device capable of generating a three-dimensional, fully enclosed potential well with zero central intensity. Through intracavity mode modulation, this fully enclosed 3D optical trap is successfully formed without requiring an external SLM, phase plates, or other complex optical systems. The advantages of this device are concentrated in the following three aspects:

First, the proposed system features a minimalist hardware structure and exceptional long-term stability. This scheme abandons the expensive and complicated SLMs commonly used to generate special-mode laser beams. Instead, it utilizes intracavity modulation to output a 532 nm TEM10 laser; by simply adjusting the positions and parameters of the extracavity components, we can readily generate spindle-shaped hollow potential wells of various dimensions. This simple hardware design not only reduces system costs and alignment complexity, but also fundamentally avoids typical issues associated with SLMs, such as parasitic interference, phase drift, and thermal damage.

Second, the system features superior 3D confinement capability. The unique topology of the optical spindle trap generates a highly enhanced gradient force. Compared to conventional hollow vortex beams, the spindle trap exhibits exceptional trap stiffness in both the axial and radial directions.

Third, the system exhibits significantly reduced phototoxicity towards biological specimens. By precisely confining target particles within the central dark-field region of extremely low light intensity, the system effectively shields living cells or bioactivator entities from high-intensity laser irradiation.

# Results

## Experimental setup

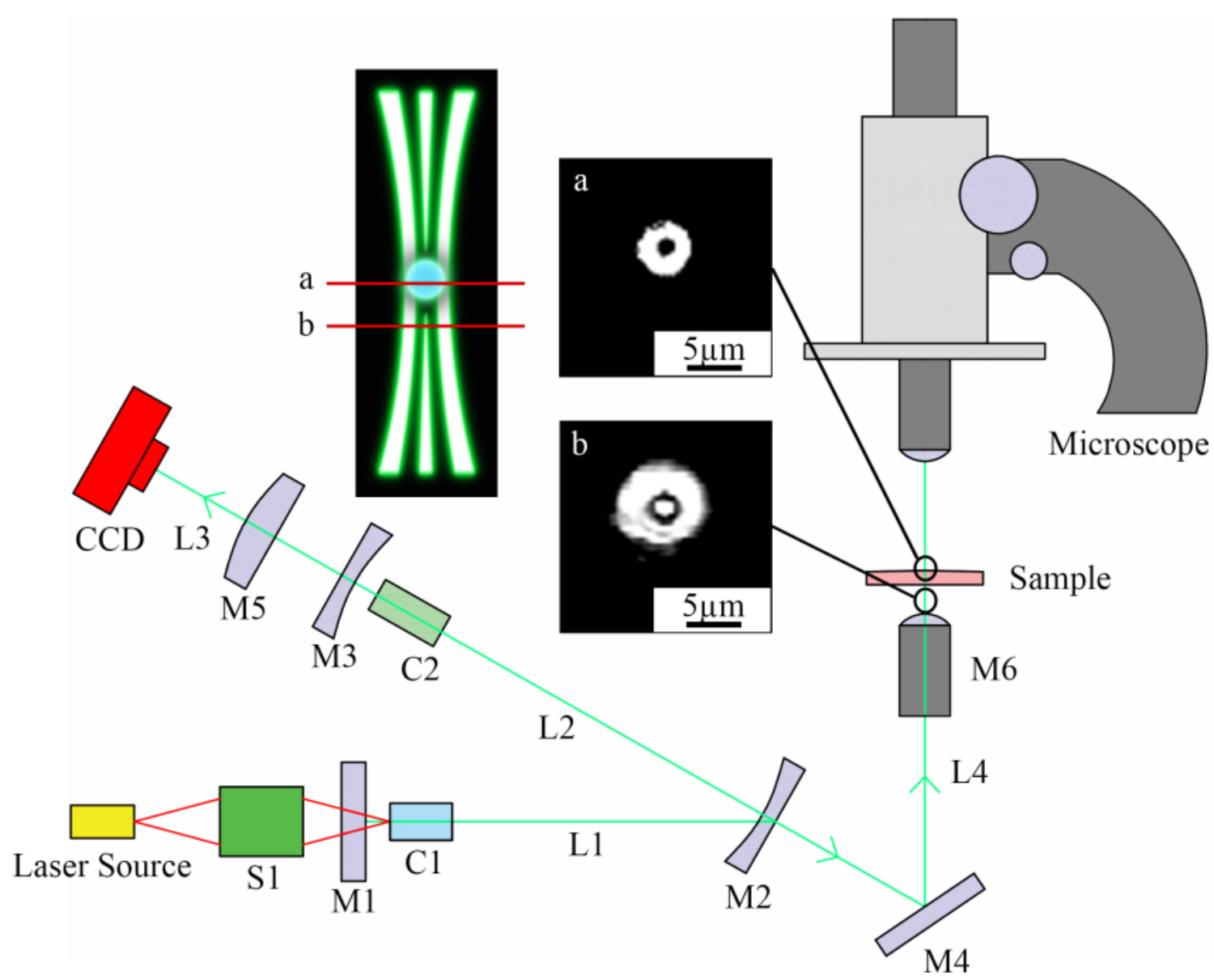

**Fig. 1.** Schematic diagram of the experimental setup. Laser source: 808 nm laser diode. S1: beam coupling optics. M1–M3: resonator mirrors, where M1 is a plane mirror with high reflectivity (HR) at 1064 nm; M2 and M3 are concave mirrors featuring HR coatings at 1064 nm and high transmission (HT) at 532 nm. M3 incorporates a microfabricated circular aperture (pinhole) with a diameter of 30 μm. C1: Nd:$YVO_4$ gain crystal; C2: LBO (lithium triborate) nonlinear crystal. M5: convex lenses. M6: objective (Plan N 10×/0.25 ∞/-/FN22). (**a-b**) Laser beam profiles at different positions. (**a**) shows the hollow beam profile at the sample plane.

The system schematic is shown in Fig. 1. The specific process of generating optical traps with 3D confinement structure mainly includes the following four steps. Firstly, the 808 nm pump source excites the gain crystal C1, establishing a fundamental Gaussian oscillation at 1064 nm. Secondly, the system converts this fundamental mode into a $TEM_{01}$ vortex beam carrying orbital angular momentum (OAM) [44] via intracavity phase modulation. Thirdly, the frequency-doubling crystal C2 converts the 1064 nm near-infrared light into 532 nm green light, simultaneously transforming the beam mode into $TEM_{10}$. Finally, the convex lenses M5 or M6 modulate the output $TEM_{10}$ mode. Then, we can construct a hollow energy potential well with 3D constraint ability near the focal plane by precisely adjusting the position and focal length of the lens.

As illustrated in Fig. 1, mirrors M1, M2, and M3 form a folded laser cavity. This configuration precisely matches the beam waists of the pump beam and the laser oscillating mode to the gain crystal and the nonlinear crystal, respectively. This precise matching maximizes pump energy conversion and improves the overall efficiency of the laser, generating L3 and L4 with identical modes. One beam is directed through an objective to generate the optical trap for particle manipulation, while the other is routed to a CCD camera for real-time monitoring of the output beam quality.

**Special three-dimensional closed optical trap**

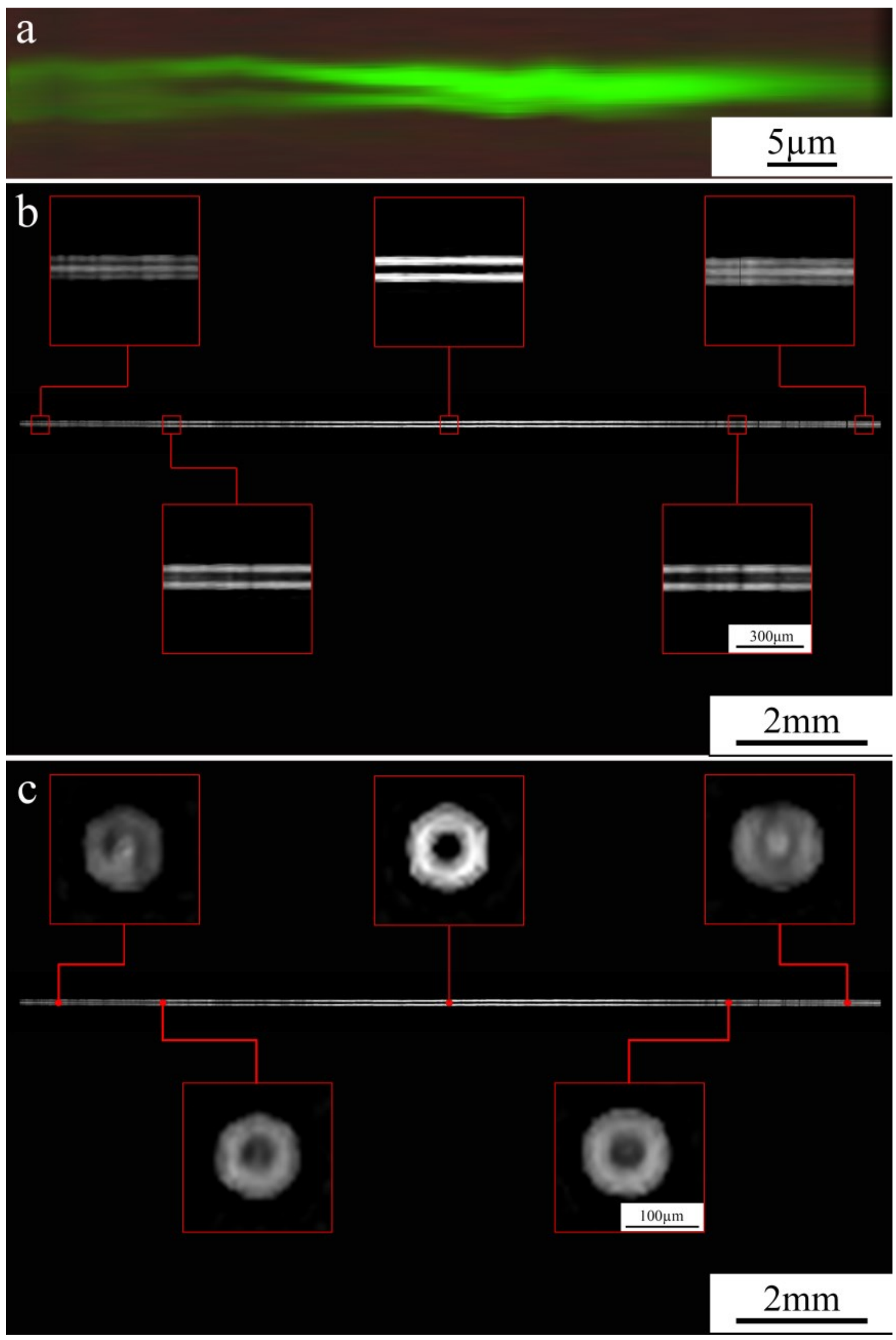

**Fig. 2.** (**a**) Optical traps used in experiments collected by microscopes. (**b-c**) Transverse beam profile captured using a convex lens (f=100 mm) positioned 250 mm from the output mirror.

Fig. 2a is reconstructed from 20 experimental light path photos collected by a microscope objective, and each photo is about 4 μm apart. Due to the large divergence angle of the M6 objective lens, the light intensity in the upper part of the optical trap (left, Fig. 2a) decreases, resulting in an asymmetrical appearance. The real situation should be a symmetrical spindle-shaped hollow structure. It is clearly observed that a spindle-shaped dark region emerges at the center of the beam. This region then shrinks and closes after a certain propagation distance, ultimately forming a fully enclosed 3D dark volume. The diameter of the spindle structure used in the experiment are about 3 μm, which approaches the detection limit of the CCD. As a result, it could only be clearly seen under a microscope.

We used a CCD to capture 850 images of the cross-section of the optical path, with an axial spacing of 20 μm between consecutive slices. We then reconstructed the 3D structure using ImageJ (Figs. 2b-c). The central energy density of the laser beam gradually decreases to approach zero within a small range and then returns to its original shape symmetrically. The length and diameter of the spindle-shaped dark region depend on the focal length of the M6 objective lens and the distance between M2 and M6 in Figure 1. Using an objective lens with a shorter focal length or increasing the distance between M2 and M6 can reduce the size of the hollow region, which can be flexibly adjusted according to the specific size of the trapped particles. In our experiment, the optical trap in the L4 experimental beam path measures approximately 25 μm in length and 3 μm in diameter, while that in the L3 observation beam path is about 6 mm long and 70 μm in diameter.

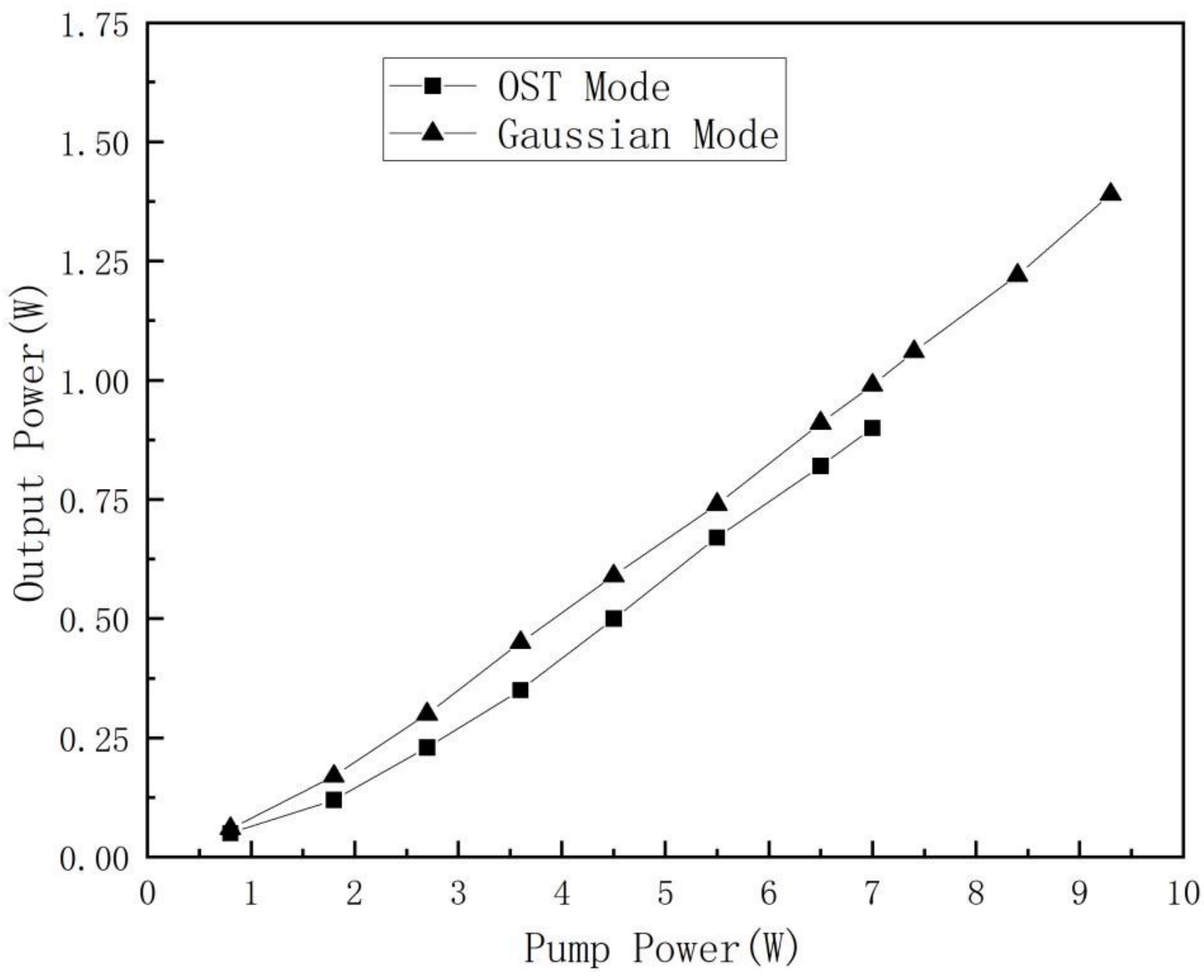

**Fig. 3.** Dependence of the resonator output power on the incident pump power.

Some studies have achieved a similar bottle-like optical structure via external phase modulation[45] or double-beam interference[46,47] (e.g., using SLMs or diffractive holograms). However, the spatial intensity profile of our hollow structure differs fundamentally from those generated by previous methods, and those approaches are typically limited by component fabrication tolerances, making it difficult to completely eliminate the residual light intensity in the bottle core. Moreover, the experimental system is complex, which leads to poor stability, low coherence, and multi-order diffraction mode interference. To solve the above problems, this work adopts a resonant cavity control strategy. By introducing nonlinear processes within the resonant cavity and combining resonance mode selection with nonlinear mode selection mechanisms, we avoid higher-order mode generation, thereby achieving highly pure single-mode output.

As shown in Fig. 3, the resonant cavity output exhibits a stable mode and linear power characteristics. The maximum output power of Gaussian mode can reach 1.5W, beyond which further increases in pump power result in saturation. The maximum output power of the $TEM_{10}$ mode is limited to approximately 1 W. Beyond this threshold, further increasing the pump power causes the beam to transition into a bad mode. The operator can flexibly adjust the mode and output power according to the nature and volume of the actually captured object. This method not only ensures the high stability of the output beam, but also shows a significant advantage in photon degeneracy, optical coherence, and quantum properties.

**Trapping of several 1 μm diameter polystyrene microspheres**

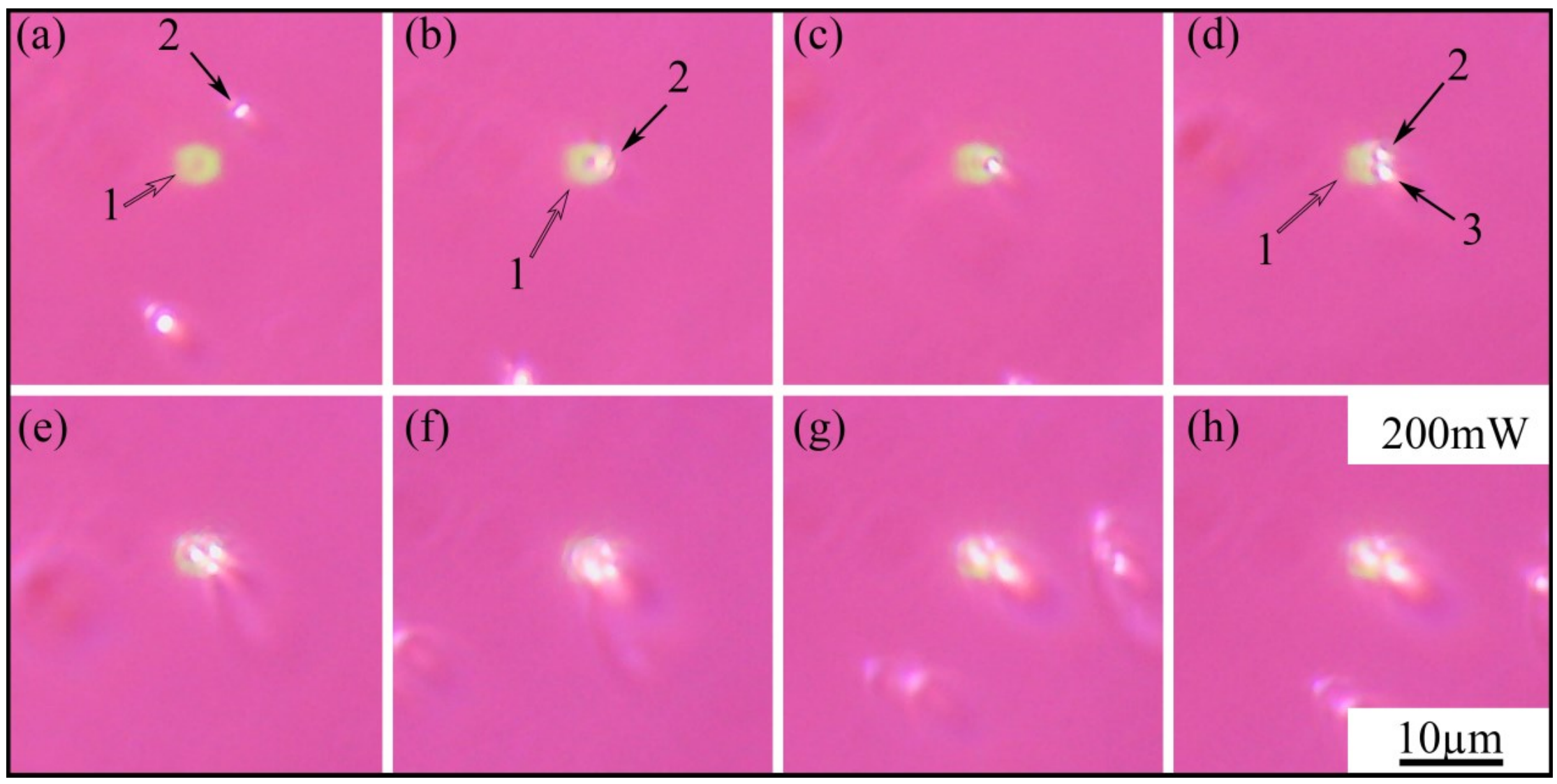

**Fig. 4.** (**a–c**) Trapping of a 1 μm diameter polystyrene microsphere using a 200 mW OST. (**d**) Laser capture of a second microsphere. (**e**) Laser capture of a third microsphere. (**f**) Laser-induced transport of the three microspheres. (**g–h**) Laser capture of additional microspheres and simultaneous manipulation of the multiple particles.

We employed a lens M6 with a short focal length and adjusted the M2–M6 distance to generate a hollow potential well with a length and diameter of 3–5 μm (Fig. 1a). Subsequently, capture experiments were performed on micron-scale particles. Figs. 4a–c illustrate the successful capture of a single, freely moving 1 μm polystyrene microsphere 2 by the 200 mW hollow laser beam 1, followed by its dragging to the left. Additionally, Figs. 4d–h depict the dynamic process in which the ring beam sweeps across the suspension, continuously collecting multiple microspheres.

**Trapping 10 μm diameter polystyrene microspheres in three-dimensional space**

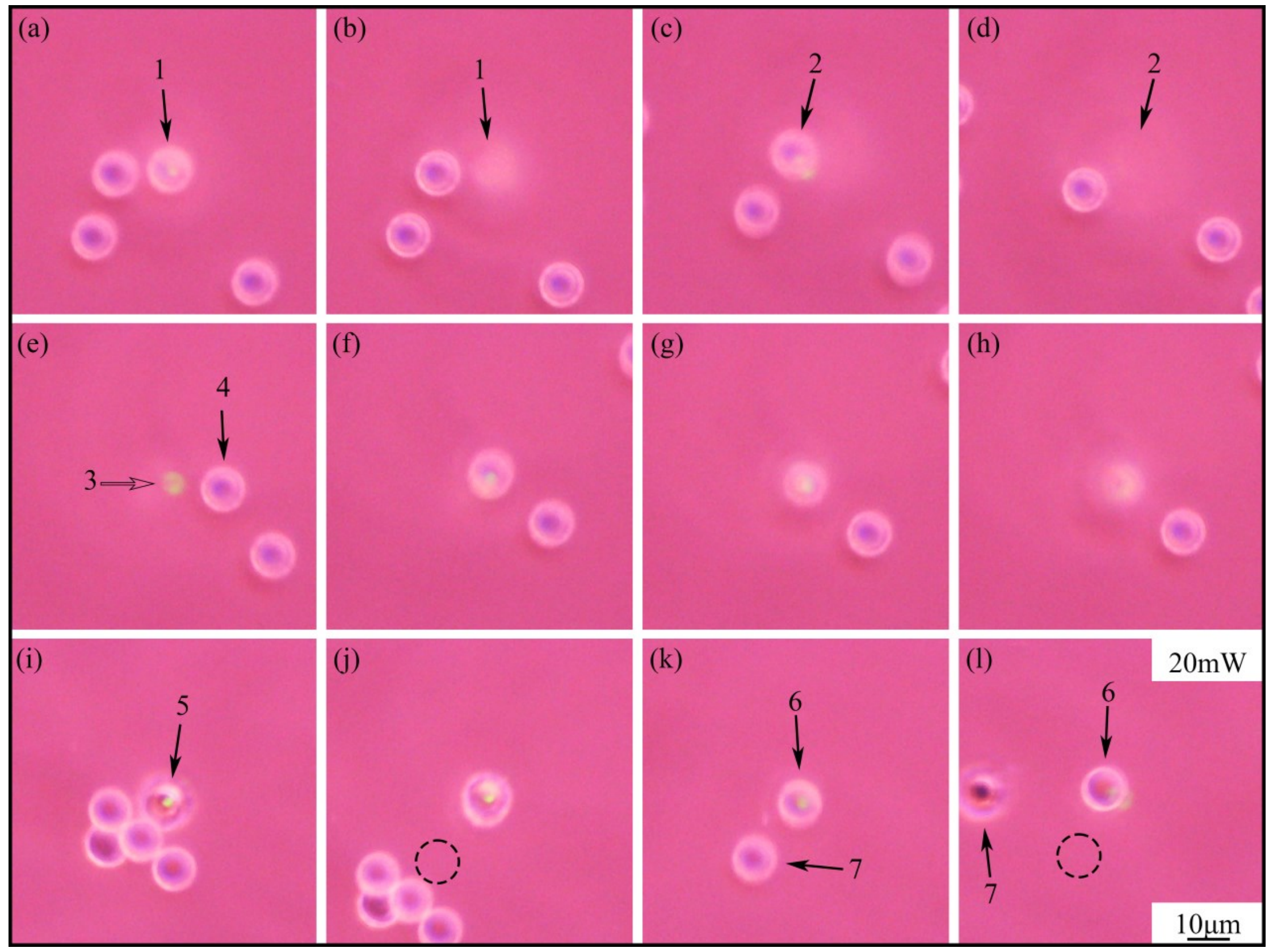

**Fig. 5.** Trapping of a 10 μm diameter polystyrene microsphere under different conditions.

Figs. 5a–d demonstrate the process of a 20 mW hollow beam capturing 10 μm polystyrene beads that are larger than the hollow diameter. As the laser beam moves towards spheres 1 and 2, the microspheres move upward and gradually move out of the focal plane of the microscope. In Figs. 5e–h, as laser beam 3 moves to microsphere 4, the entire process of the sphere's slow upward movement is observed. This demonstrates the axial trapping capability of the optical tweezers, with the microsphere eventually being stably trapped within the central dark core of the hollow beam in 3D space. Figs. 5i–j show the microspheres trapped at an elevated position, where the laser beam moves at this height to drag sphere 5. In Figs. 5k–l, the laser beam drags microsphere 6, while microsphere 7 serves as a stationary reference.

## Trapping mouse liver cells

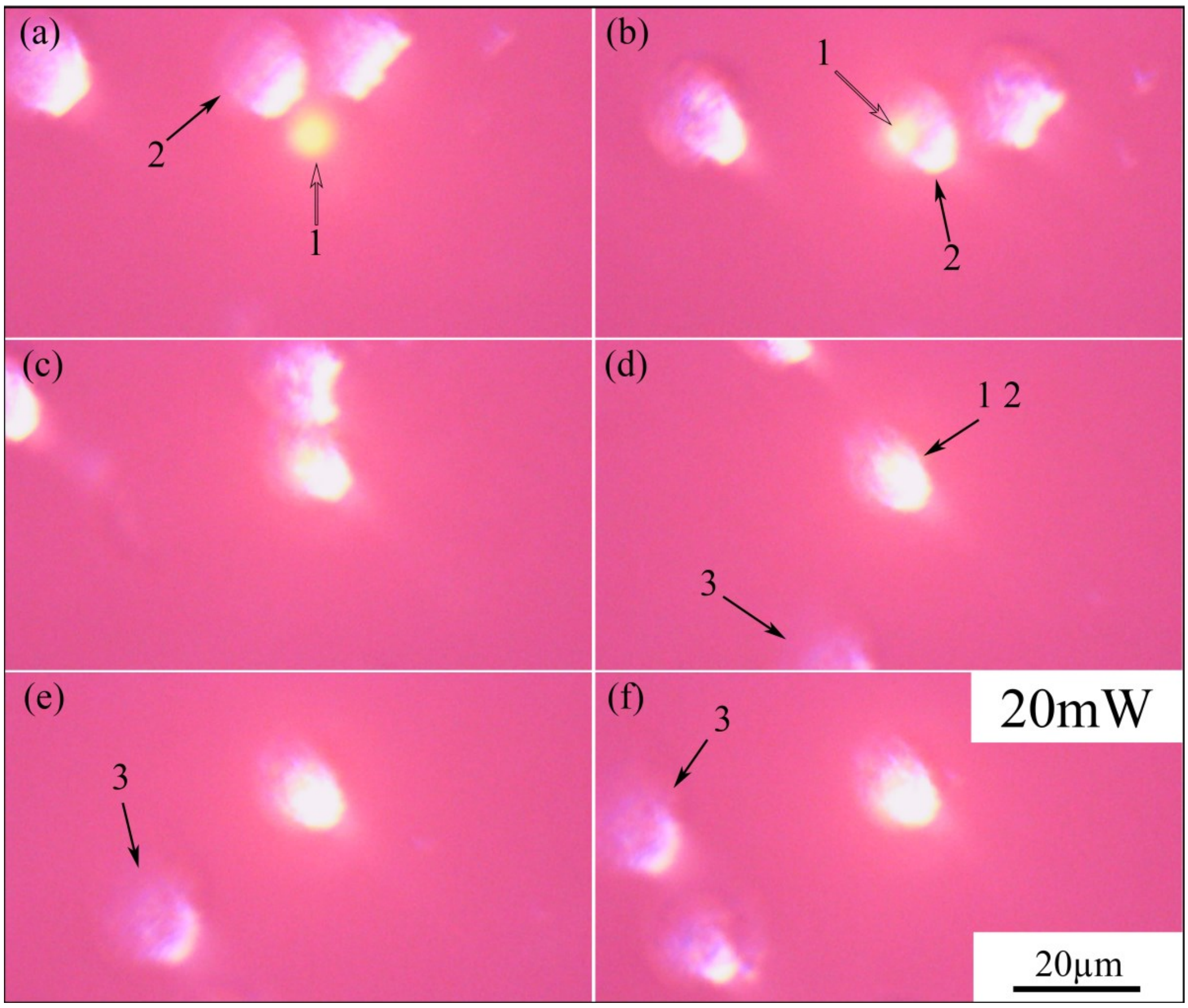

**Fig. 6.** Dynamic optical trapping and 3D manipulation of a biological cell using the structured hollow beam. (**a**) Pre-trapping configuration showing the hollow beam 1 and the target cell 2. (**b**) Successful confinement of cell 2 within the central intensity null. (**c–f**) Controlled lateral translation of the trapped cell, with an adjacent unmanipulated cell 3 serving as a stationary biological reference.

Fig. 6 demonstrates the high-speed, stable capture of mouse hepatocytes at a low optical power of 20 mW. This trap's low power requirement and hollow structure with minimal intensity in the center allow the cells to survive for a longer time.

## Discussion

We developed a 3D, fully enclosed OST featuring a central dark core with near-zero intensity. Using this trap, we successfully confined microspheres and biological cells of various sizes. We find that the length and diameter of the OST are related to the focal length of the objective lens and the distance between the objective lens and the M2. By tuning these two parameters, the dimensions of the OST can be flexibly adjusted. This mechanism provides a cost-effective and exceptionally straightforward approach for precisely controlling the geometry and trap stiffness of the 3D optical trap. The system shows high internal physical stability in long-term continuous operation, which provides a reliable guarantee for experiments that need long-term stable use of optical traps. Experimental results demonstrate that this optical trap can stably confine single or multiple micrometer-scale particles, effectively suppressing particle escape and enabling long-term trapping. This dark-field trapping mechanism minimizes localized heating and phototoxicity to biological samples, making it ideally suited for the long-term, non-destructive observation and manipulation of living specimens. Although there is no direct comparison test temporarily, its hollow structure solves the shortcoming that Gaussian optical tweezers bind cells at the waist, which easily leads to thermal damage or photobleaching.

In the future, we will further upgrade the observation system and conduct quantitative comparative experiments between the proposed system and Gaussian optical tweezers. By incorporating a high-precision, height-adjustable mount for the M6 objective, we will achieve real-time dynamic tuning of the optical trap dimensions, enabling the trap to flexibly accommodate particles of various sizes. Furthermore, this three-dimensional hollow structure is expected to overcome the limitations of conventional optical tweezers, enabling the stable trapping of absorbing particles (such as metallic nanoparticles or pigment cells).

In summary, these results show that this 3D fully enclosed optical trap holds potential in the trapping and manipulation of particles. This technology provides a more moderate and efficient manipulation platform for the research of micro and sub-micron scales (especially sensitive samples such as biological cells). Therefore, this OST has potential to surpass Gaussian and vortex optical tweezers in the fields of physics, chemistry, biology, and materials science.

# Methods

**OST setup**

The pump light source is a semiconductor laser of FOCUSLIGHT (FL-DLS02-FCSB04-I-30-808-10), with a maximum output power of 30W, a laser spot diameter of 200 μm and an output wavelength of 808nm. The parameter of laser output coupling mirror S1 is 1: 0.8. In the laser resonator, the curvature radius of the concave mirror M2 is 200 mm, and the curvature radius of the concave mirror M3 is 100 mm. M1 is a plane mirror with high reflectivity (HR) at 1064 nm; M2 and M3 are concave mirrors featuring HR coatings at 1064 nm and high transmission (HT) at 532 nm. M3 incorporates a microfabricated circular aperture (pinhole) with a diameter of 30 μm. The distances between S1 and M1 are 40mm, M1 and C1 are 10mm, C1 and M2 are 160mm, M2 and C2 are 90mm, C2 and M3 are 70mm, and M2 and M6 are 900mm. C1 is an Nd:YVO4 crystal (3 × 3 × 10 mm) cooled by a water chiller (LX-150, 150 W, 20 °C), while C2 is an LBO crystal (2 × 2 × 12 mm) regulated by a TEC module to maintain its optimal operating temperature of 30 °C. A convex lens with a 100 mm focal length is used as M5 in the observation optical path. The distance between M3 and M5 is set to 250 mm, resulting in a hollow spot size suitable for CCD observation. M6 on the experimental optical path uses the objective lens of Olympus(Plan N 10×/0.25 ∞/-/FN22). The microscope uses a 20-fold objective lens with an electronic eyepiece with 1.8 million pixels and a 25-fold magnification. The CCD sensor features a resolution of 720 × 576 pixels with a physical pixel size of 5 × 6.25 μm, enabling a measurement accuracy of 6.25 μm. Capable of detecting spot sizes ranging from a minimum of 5 × 6.25 μm to a maximum of 3595 × 3595 μm, this setup allows for the direct visualization of the spot size, shape and intensity distribution, facilitating design adjustments to optimize beam quality. We use a 300-μm-thick quartz slide for the sample cell.

**Preparation of polystyrene microsphere samples**

Polystyrene microsphere powder (1 μm or 10 μm in diameter) is dispersed in deionized water to achieve a concentration suitable for microscopic observation. After thorough shaking, a droplet of the suspension is deposited onto the quartz slide.

## Preparation of cell samples

We suspend the cells in DMEM to achieve a concentration suitable for microscopic observation, and subsequently deposit a droplet onto the quartz slide.

## Acknowledgements

We are grateful to Professor Yunhong Zhang (School of Chemistry and Chemical Engineering, Beijing Institute of Technology) and Dr Jiuyi Sun (School of Chemistry and Chemical Engineering, Beijing Institute of Technology) for helpful discussions on equipment construction. We are grateful to the Analysis & Testing Center (Beijing Institute of Technology) for providing the necessary space and facilities.

## Author Contributions

C.Z. conceived and designed the study. C.Z. and X.F. constructed the experimental setup and performed the experiments. X.F. wrote the manuscript. C.Z. helped with manuscript revision. C.Z. oversaw and directed the whole project. P.M. provided experimental sites. D.Z. provided experimental cells. All authors read and approved the final manuscript.

## Competing Interests Statement

The authors declare no competing interests.